\documentclass[english]{article}%
\usepackage[T1]{fontenc}
\usepackage[latin1]{inputenc}
\usepackage{amsmath}
\usepackage{babel}
\usepackage{graphicx}
\usepackage{amsfonts}
\usepackage{amssymb}%
\makeatletter
\newtheorem{theorem}{Theorem}
\newtheorem{acknowledgement}[theorem]{Acknowledgement}

\makeatother
\begin{document}

\title{{\large Four-level systems and a universal quantum gate}}
\author{M.C. Baldiotti\thanks{E-mail: baldiott@fma.if.usp.br} and D.M.
Gitman\thanks{E-mail: gitman@dfn.if.usp.br}\\Instituto de Física, Universidade de São Paulo,\\Caixa Postal 66318-CEP, 05315-970 São Paulo, S.P., Brazil}
\maketitle

\begin{abstract}
We discuss the possibility of implementing a universal quantum XOR gate by
using two coupled quantum dots subject to external magnetic fields that are
parallel and slightly different. We consider this system in two different
field configurations. In the first case, parallel external fields with the
intensity difference at each spin being proportional to the time-dependent
interaction between the spins. A general exact solution describing this system
is presented and analyzed to adjust field parameters. Then we consider
parallel fields with intensity difference at each spin being constant and the
interaction between the spins switching on and off adiabatically. In both
cases we adjust characteristics of the external fields (their intensities and
duration) in order to have the parallel pulse adequate for constructing the
XOR gate. In order to provide a complete theoretical description of all the
cases, we derive relations between the spin interaction, the inter-dot
distance, and the external field.

\end{abstract}

\section{Introduction}

At present there exists a belief that single quantum systems, e.g., atomic
traps, and electronic devices whose operation involves only a few number of
electrons, could be used to implement the so-called quantum gates for quantum
computations. Quantum computations (quantum computers, or simply QC in what
follows) can efficiently solve problems that are considered intractable by
classical computers, e.g., the factoring of primes \cite{Shor94} and the
simulation of others quantum systems \cite{Fey82}. As for classical computers,
for the QC the accomplishment of an arbitrary algorithm can be performed using
just a few specific manipulations called universal quantum gates. With these
universal quantum gates, a process that acts in an arbitrary number of quantum
bits (qubits) can be constructed using gates that act only in one and two qubits.

Although the first realizations of quantum computers were performed using
trapped ions \cite{MonMeKIW95}, NMR \cite{ChuGerK98}, and optical cavities
\cite{TurHoLMK95}, it is believed that, as for classical computers, for QC the
most promising object for a possible large-scale implementation, i.e., a
implementation involving a large number of qubits, will be the solid-state
devices. Among these devices, one can highlight the system of two coupled
semiconductor quantum dots (QD) \cite{LosDi98}. In this system the qubits are
the one-half spin states of an excess electron in each dot, and the universal
quantum gates can be obtained by performing arbitrary rotation of the
individual spin (one-qubit gate) and one operation capable of entangling the
spins of the two electrons (two-qubits gate). Any such operation with this
entangling feature can be used \cite{BreDaDGHMNO02}. The rotations and the
entanglement operations can be performed by applying external electromagnetic
pulses to the dots. These pulses can be applied in a serial manner, which
perform sequentially the one and the two-qubit gates using a sequence of
pulses, or in a parallel manner, where the two and one-qubit gates are
performed at once by the application of just one adequate pulse (called a
\textit{parallel pulse}). Due to the decohering effects it is important to
perform the operations in a minimal time, which, as discussed in
\cite{BurLoDS99}, can be achieved more adequately with a parallel pulse.

The adequate parallel pulse to implement a quantum gate can be obtained if we
know the Hamiltonian that describes the coupled QD, but to construct this
Hamiltonian, one has to know the exact relation between the external
parameters of the system (such as the external fields and the physical
characteristics of the dot) and the interaction between the electrons. An
advance in the description of this relation was obtained in \cite{BurLoD99},
where the authors have used the Heitler-London approximation to obtain an
explicit expression that relates the interaction between the dots, the
inter-dots distance and the external electromagnetic field. However, in this
article, the two dots were subject to the same magnetic field, and, as we will
see, a little difference between the fields in the dots possess a great
influence in the evolution of the system and, consequently, in the
implementation of the gates. In addition, after the Hamiltonian is known, we
have to construct exact solutions of the evolution equation of this four-level
system. In our work \cite{BagBaGL07} we studied possible exact solutions of a
system of two coupled spin one-half particles subject to external
time-dependent magnetic fields.

In the present work we discuss one possibility of implementing an universal
quantum gate, namely the \textit{exclusive or} (XOR) gate, by using two
coupled QD subjected to external magnetic fields that are parallel and
slightly different. First, in Sections \ref{4LS}\ and \ref{PF}, we make a
brief discussion of the model with the parallel external magnetic fields of
general form, and describe possible corresponding exact solutions from our
previous work \cite{BagBaGL07}. In order to have a complete theoretical
description of the problem, one needs to find relations between the spin
interaction and external fields. This problem is solved in Section \ref{HL}.
To this end we generalize results of \cite{BurLoD99}, derived for equal
parallel external fields, to the case of different fields, obtaining relations
between the spin interaction, the inter-dot distance, and the external field.
In section \ref{XOR} we consider two parallel fields with specific form. In
the first case, parallel external fields with intensity difference at each
spin being proportional to the interaction between the spins, the latter
depending on time. General exact solution describing this system is presented
and analyzed to adjust field parameters. Then we consider parallel fields with
intensity difference at each spin being constant and interaction between the
spins switching on and off adiabatically. In both cases we adjust
characteristics of the external fields (their intensities and duration) in
order to have the parallel pulse adequate for constructing the XOR gate.

\section{Four-level system\label{4LS}}

In non relativistic quantum mechanics the dynamics of a fixed spin one-half
particle, subject to a time-dependent external field $\mathbf{K}\left(
t\right)  $, can be described by the Hamiltonian $\hat{h}=\left(
\boldsymbol{\sigma} \cdot\mathbf{K}\right)  $ \cite{BagGiBL05}, where
$\boldsymbol{\sigma} =\left(  \sigma_{1},\sigma_{2},\sigma_{3}\right)  $ are
the Pauli matrices and we set $\hbar=1$. The quantum Hamiltonian of two
interacting one-half spins subjected to external fields $\mathbf{G}$ and
$\mathbf{F}$, respectively, is chosen as, see e.g. \cite{Tow92},%
\begin{equation}
\hat{H}\left(  \mathbf{G,F,}J\right)  =\left(  \boldsymbol{\rho} \mathbf{\cdot
G}\right)  +\left(  \boldsymbol{\Sigma} \mathbf{\cdot F}\right)  +\frac{J}%
{2}\left(  \boldsymbol{\Sigma} \mathbf{\cdot}\boldsymbol{\rho} \right)  \,,
\label{1}%
\end{equation}
where the function $J=J\left(  t\right)  $ defines the spin interaction,
$\mathbf{G}=\left(  G_{1}\left(  t\right)  ,G_{2}\left(  t\right)
,G_{3}\left(  t\right)  \right)  $ and $\mathbf{F}=\left(  F_{1}\left(
t\right)  ,F_{2}\left(  t\right)  ,F_{3}\left(  t\right)  \right)  $ are the
time-dependent external fields at each particle, and the $4\times
4$\ Dirac-matrices $\boldsymbol{\rho} $ and $\boldsymbol{\Sigma} $ are in the
standard representation
\begin{equation}
\boldsymbol{\Sigma} =I\otimes\boldsymbol{\sigma} \,,\;\boldsymbol{\rho}
=\boldsymbol{\sigma} \otimes I\,,\left(  \boldsymbol{\Sigma} \cdot
\boldsymbol{\rho} \right)  =\boldsymbol{\sigma} \otimes\boldsymbol{\sigma}
=\sum_{i=1}^{3}\sigma_{i}\otimes\sigma_{i}\,, \label{eq3}%
\end{equation}
where $I$ is the $2\times2$ identity.

The interaction of the spins in (\ref{1}) is known as the \textit{Heisenberg
interaction} and describes, for example, the interactions in the well known
Hubbard model \cite{AshMe76}. In particular, under certain conditions, this
interaction can be used to describe the coupling between two QD \cite{LosDi98}%
. In this case, it is possible to control not only the external fields, but
also the spin interaction. When these QD are used to implement a quantum gate,
the operation of the gate is performed by varying the external fields and the
spin interaction during a certain time $\tau$. To this end, in order to
justify the Heisenberg interaction, the following conditions (see
\cite{LosDi98}) must hold:

\begin{enumerate}
\item The time $\tau$ cannot be too small to avoid transitions to higher
energy levels, so that the time scale should be bigger than $\hbar/\Delta E$,
where $\Delta E$ is the difference between the first two\ energy levels;

\item The decoherence time of the physical system should be much bigger than
$\tau$.
\end{enumerate}

The use of these QD to implement quantum algorithms requires controlling the
individual spins, which can be done by the fields $\mathbf{F}$ and
$\mathbf{G}$, and any interaction between the spins capable of creating a
entangled state starting from a original product state \cite{BreDaDGHMNO02}. A
system described by the Hamiltonian (\ref{1}) satisfies these conditions.
Namely, when the fields $\mathbf{F}$ and $\mathbf{G}$ are zero, the evolution
operation $R_{t}\left(  \mathbf{G,F,}J\right)  $ of the Hamiltonian (\ref{1})
can be written as \cite{BagBaGL07}%
\begin{align}
&  R_{t}\left(  0,0\mathbf{,}J\right)  =\exp\left[  i\Phi\left(  t\right)
/2\right]  \left[  \mathbb{I}\cos\Phi\left(  t\right)  -iA\sin\Phi\left(
t\right)  \right]  \,,\nonumber\\
&  A=\frac{1}{2}\left[  \mathbb{I}+\left(  \boldsymbol{\Sigma}\mathbf{\cdot
}\boldsymbol{\rho}\right)  \right]  ~,\ \Phi\left(  t\right)  =\int_{t_{0}%
}^{t}J\left(  \tau\right)  \,d\tau\,, \label{3}%
\end{align}
where $\mathbb{I}$ is the $4\times4$ unity matrix. From the above expression
we see that, when $\Phi=\pi/4$, the evolution operator acts as the universal
gate known as square root of swap ($U_{\mathrm{sw}}^{1/2}$). The interaction
function $J$ can experimentally be controlled in many different ways, e.g., by
applying electrical or magnetic fields to the dots \cite{PetJoTLYLMHG05}. So
we can construct any quantum gate by a sequence of pulses by turning on and
off the external fields and the interaction. For example, the XOR gate can be
constructed as \cite{BurLoD99}%
\begin{equation}
U_{\mathrm{XOR}}=\exp\left(  i\frac{\pi}{4}\rho_{3}\right)  \exp\left(
-i\frac{\pi}{4}\Sigma_{3}\right)  U_{\mathrm{sw}}^{1/2}\exp\left(  i\frac{\pi
}{2}\rho_{3}\right)  U_{\mathrm{sw}}^{1/2}~. \label{4}%
\end{equation}
On the other hand, the construction of gates by a sequence of pulses is not
appropriate, because the duration of all the sequence can be too long,
violating condition (2) above, or the pulses need to vary too fast, violating
the condition (1). So it is important to be able to implement the gates at
once, applying just one adequate field. This single pulse is called parallel
pulse \cite{BurLoDS99}. For example, in the case of the XOR gate (\ref{4}), to
use a parallel pulse one needs to find a field whose evolution operation, at a
given instant $\tau$ of time, has the form%
\begin{equation}
R_{\tau}\left(  \mathbf{G,F,}J\right)  =U_{\mathrm{XOR}}=\exp\left[
-i\frac{\pi}{4}\left(  \Sigma_{3}\rho_{3}+\Sigma_{3}+\rho_{3}\right)  \right]
~. \label{4a}%
\end{equation}
In order to find this parallel pulse, in a general case, we need to know exact
solutions of the Schrödinger equation with the Hamiltonian (\ref{1}) for
different kinds of external fields and interactions. A large number of these
exact solutions are present in our previous work \cite{BagBaGL07}, and here we
will use some of these results to describe the implementation of a
$U_{\mathrm{XOR}}$ gate. The procedure developed here can easily be extended
to others gates using others fields present in \cite{BagBaGL07}.

\section{Parallel external fields\label{PF}}

Let each spin in our system be subject to different time-dependent external
magnetic fields along the $z$ direction,
\begin{equation}
\mathbf{G}=\left(  0,0,\mu_{B}g_{1}B_{1}\right)  \,,\;\mathbf{F}=\left(
0,0,\mu_{B}g_{2}B_{2}\right)  ~,\;B_{1,2}=B_{1,2}\left(  t\right)
\mathrm{\,}. \label{5}%
\end{equation}
where $\mu_{B}$ is the Bohr magneton and $g_{i}$ the $g$-factor of the dot
$i$. Therefore the Hamiltonian (\ref{1}) assumes the form
\begin{equation}
\hat{H}=\frac{1}{2}\left[  \left(  \Sigma_{3}+\rho_{3}\right)  B_{+}-\left(
\Sigma_{3}-\rho_{3}\right)  B_{-}-J\right]  +AJ~,\;B_{\pm}=\mu_{B}\left(
g_{1}B_{1}\pm g_{2}B_{2}\right)  \,, \label{6}%
\end{equation}
with the constant $4\times4$ orthogonal matrix $A$ given in (\ref{3}). For the
first and fourth components $v_{1}$ and $v_{4}$ of the four-spinor $\Psi$,
solution of the Schrödinger equation ($i\dot{\Psi}=\hat{H}\Psi$) with the
above Hamiltonian, we get%
\begin{equation}
v_{1}=C_{1}\exp\left[  -i\int_{0}^{t}\left(  \frac{J}{2}+B_{+}\right)
\,d\tau\right]  \,,\;v_{4}=C_{4}\exp\left[  -i\int_{0}^{t}\left(  \frac{J}%
{2}-B_{+}\right)  \,d\tau\right]  \,, \label{3f}%
\end{equation}
where $C_{1,4}$ are complex constants. Besides, for the components $v_{2,3}$
of $\Psi$, we obtain the expression%
\begin{align}
&  i\dot{\psi}^{\prime}=\left[  \left(  \boldsymbol{\sigma}\mathbf{\cdot
K}\right)  -\frac{J}{2}\right]  \psi^{\prime}\,,\;\psi^{\prime}=\left(
\begin{array}
[c]{c}%
v_{2}\\
v_{3}%
\end{array}
\right)  \,,\label{3e}\\
&  \mathbf{K}\left(  t\right)  =\left(  J\left(  t\right)  ,0,B_{-}\left(
t\right)  \right)  \,. \label{3b}%
\end{align}
Making the transformation%
\begin{equation}
\psi^{\prime}\left(  t\right)  =\exp\left[  \frac{i}{2}\int_{0}^{t}J\left(
\tau\right)  \,d\tau\right]  \psi\left(  t\right)  \,, \label{3g}%
\end{equation}
the two-component spinor $\psi$ obeys the equation for a single particle with
spin one-half in an effective external field $\mathbf{K}\left(  t\right)  $,
i.e., $i\dot{\psi}=\left(  \boldsymbol{\sigma}\mathbf{\cdot K}\right)  \psi$.
Consequently, in this case, the four-level system problem reduces to finding
solutions of a two-level system. Writing the solution of this two-level-system
as $\psi\left(  t\right)  =\hat{u}_{t}\psi\left(  0\right)  $, using the
$2\times2$ evolution operator $\hat{u}_{t}\left(  J,B_{-}\right)  $, we can
write the evolution operator of the four-level-system governed by the
Hamiltonian (\ref{6}) as%
\begin{align}
&  R_{t}\left(  \mathbf{G,F,}J\right)  =\exp\left(  -\frac{i}{2}\left[
\left(  \Sigma_{3}+\rho_{3}\right)  \Gamma\left(  t\right)  +\Sigma_{3}%
\rho_{3}\Phi\left(  t\right)  \right]  \right)  M\left(  t\right)
~,\nonumber\\
&  \Gamma\left(  t\right)  =\int_{0}^{t}B_{+}\left(  \tau\right)
\,d\tau~,\ \Phi\left(  t\right)  =\int_{0}^{t}J\left(  t\right)  \,d\tau~,
\label{4.1}%
\end{align}
with the $4\times4$\ matrix $M\left(  t\right)  $ given by components of the
single particle evolution operator $\hat{u}_{t}$ as%
\[
M=\left(
\begin{array}
[c]{ccc}%
1 & 0 & 0\\
0 & \hat{u}_{t} & 0\\
0 & 0 & 1
\end{array}
\right)  ~.
\]
Writing the effective field $\mathbf{K}$ (\ref{3b})\ as%
\begin{equation}
\mathbf{K}=\frac{\mathbf{\dot{q}+}\left[  \mathbf{q\times\dot{q}}\right]
}{1+\mathbf{q}^{2}}~,\;\dot{q}_{2}=q_{1}\dot{q}_{3}-q_{3}\dot{q}_{1}~,
\label{8.6}%
\end{equation}
for an appropriate real three-vector $\mathbf{q}\left(  t\right)  $, we can
obtain the evolution operator $\hat{u}_{t}$ in the general form
\cite{BagGiBL05}%
\begin{equation}
\hat{u}_{t}=\frac{1+\mathbf{qq}_{0}-i\mathbf{\sigma p}}{\sqrt{\left(
1+\mathbf{q}^{2}\right)  \left(  1+\mathbf{q}_{0}^{2}\right)  }}%
\,,~\mathbf{p}=\mathbf{q-q}_{0}+\left[  \mathbf{q}_{0}\times\mathbf{q}\right]
~, \label{8.8a}%
\end{equation}
where $\mathbf{q}_{0}=\mathbf{q}\left(  0\right)  $. Thus we can construct the
XOR operator (\ref{4a}) by using any effective field $\mathbf{K}$ given by a
vector $\mathbf{q}$ that, after a certain time $T$, points again in the
initial direction ($\mathbf{q}\left(  T\right)  \propto\mathbf{q}_{0}$). When
this is the case, $\hat{u}_{T}=I$ and the evolution operator $R_{t}$
(\ref{4.1})\ will assume the form $U_{\mathrm{XOR}}$ (\ref{4a})\ when%
\[
\Gamma\left(  T\right)  =\Phi\left(  T\right)  =\frac{\pi}{2}%
\operatorname{mod}\left(  2\pi\right)  ~.
\]
This periodicity in the direction of the vector $\mathbf{q}$ can be considered
a general property for which the parallel field acts as a XOR gate. An
arbitrary choice of this vector, respecting only this periodic condition, can
be used to obtain a variety of parallel fields that will certainly possess the
adequate characteristics.

Although the interaction function $J$ depends on the applied external fields,
as we will see in the next section, when $B_{-}<<B_{+}$, we can make
$J=J\left(  B_{+}\right)  $ and consider the vector $\mathbf{K}$ in (\ref{3b})
as composed of two independent functions $J\left(  t\right)  $ and
$B_{-}\left(  t\right)  $. Besides, the interaction function can be controlled
by electric fields \cite{PetJoTLYLMHG05}, whose interference in the spin
states via the spin-orbit coupling, in many practical application, can be
neglected. Then, if necessary, we can consider the functions $J,~B_{-}$ and
$B_{+}$ independently. In addition, although the obtained expressions depend
on the sum and on the difference between the fields at the spins, only the
average value of this sum is relevant, so that the explicit form of its time
variation can be arbitrary.

Some comments on the experimental realization are in order. With the
technology available nowadays, systems of two coupled QD, each holding exactly
one electron, are routinely constructed in experimental solid state physics
\cite{AwsLoS02}. The confinement of the electron and the control of the
interaction between them can be performed by electrical gates directly
assembled in the semiconductor material using etching techniques. The coupled
electron spin is a result of the combination of the Coulomb interaction and
the Pauli Exclusion Principle. The spin state of this system can be
initialized using a variety of techniques \cite{Ata06,PetJoTLYLMHG05}. In all
experiments with such systems, the difference between the energy of singlet
and triplet spin states are controlled by applying either static or
time-dependent external magnetic fields of intensity up to the order of a few
teslas. See, for example, the experimental setup \cite{PetJoTLYLMHG05}.
Therefore, the construction of systems of two-coupled QD subject to external
parallel magnetic fields is commonplace. However, the main problem in the
experimental application of the results presented in this work is to generate,
and control, a difference in the magnetic field felt by each spin, that is,
the manipulation of $B_{-}$\ (\ref{6}). The quantity $B_{-}$ is not
necessarily associated with a gradient in the applied magnetic field. It is
possible to couple two different QD \cite{BraScDSPKWRG06} in such a way that
$g_{1}\neq g_{2}$. In addition, some techniques permit the manipulation of the
$g$-factor by changing the size of the dots or by the application of external
electromagnetic fields \cite{MedRiW03,DesSe06}. The hyperfine coupling between
the electron spin and the nuclear spins of the semiconductor material can also
be explored to obtain a field gradient by producing a differential Overhauser
field \cite{JohPeTYLMHG05}. As a result, we can have a nonzero $B_{-}$ even
when the two dots are subject to the same external magnetic field. Besides, it
is possible to fine-tune the quantity $B_{-}$ with the application of a
localized magnetic field \cite{GolLo02}. In \cite{KopBuTVNMKV06} an
experimental setup of this kind is used to control the spin state of a single
electron in a coupled QD system. The fine-tuning of the quantity $B_{-}$ has
been an intense object of study in the current solid state experiments.

\section{Relation between the interaction function and the external
fields\label{HL}}

In order to construct the evolution operator (\ref{4.1}) for some specific
parallel field, one needs to know the explicit dependence between the external
fields and the interaction function. In the article \cite{BurLoD99}, the
authors study two QD, coupled with the Heisenberg interaction (\ref{1}), and
use the Heitler-London approximation to obtain an expression for the
interaction function $J\left(  V,a,B,E\right)  $ as a function of the gate
voltage between the coupled QD ($V$), the inter-dot distance ($a$), an
external magnetic field ($B$) and an external electric field ($E$). The
knowledge of the interaction as a function of various parameters allows to
compensate the changes in $J$ with the variation of one given parameter by
controlling another parameter. So one can, for example, vary the magnetic
field $B$ and maintain $J$ constant by changing the inter-dot potential $V$.
This procedure allows treating some parameters of the Hamiltonian of the
system as independent. However, the analysis developed in \cite{BurLoD99}
concerns only the problem of two dots subject to the same magnetic field.
Since the work developed here depends on the difference between the fields at
each dot, we will repeat the process given in the cited article, but with
different magnetic fields at each dot. For different parallel\ magnetic fields
$\left(  B_{1},B_{2}\right)  $\ applied at the spins, the non-homogeneity of
the field make it very complicated to calculate $J\left(  B_{1},B_{2}\right)
$, since, in this case, the Zeeman terms ($\boldsymbol{\rho}\mathbf{\cdot G}$
and $\boldsymbol{\Sigma}\mathbf{\cdot F}$) will not be independent of the
space coordinates anymore. Despite this fact, since only a small field
difference is important, the fields can be considered homogeneous inside each
dot and the Heitler-London approximation can still be used to describe the
interaction function.

The model consists of two identical bi-dimensional dots with an inter-dot
separation of $2a$, each dot with one electron of charge $e$ and subject,
respectively, to the magnetic field $B_{1}$ and $B_{2}$ in the $z$ direction.
The field $B_{i}$ is considered homogeneous inside the dot $i$. In the
Heitler-London approximation, we start with a combination of the orbital
ground states of bi-dimensional single dots and combine this state in
symmetric ($\left\vert \Psi_{+}\right\rangle $) and anti-symmetric
($\left\vert \Psi_{-}\right\rangle $) states for the double dot problem as%
\[
\left\vert \Psi_{\pm}\right\rangle =\frac{\left\vert -+\right\rangle
\pm\left\vert +-\right\rangle }{\sqrt{2\pm S^{2}}}%
\]
where $\varphi_{\pm}\left(  \mathbf{r}\right)  =\left\langle \mathbf{r}%
\right.  \left\vert \pm\right\rangle $ denotes the one-particle orbital
centered at $\mathbf{r}=\left(  \pm a,0\right)  $ and $S=\left\langle
\mathbf{+}\right.  \left\vert -\right\rangle $ the overlap between the two
orbitals. The interaction function, or the exchange energy, is the difference
between the energies, $J=\left\langle \Psi_{-}\right\vert H_{orb}\left\vert
\Psi_{-}\right\rangle -\left\langle \Psi_{+}\right\vert H_{orb}\left\vert
\Psi_{+}\right\rangle $ where the double-dot orbital Hamiltonian is given by%
\begin{align}
&  H_{orb}=h^{0}\left(  \mathbf{r}_{1}\right)  +h^{0}\left(  \mathbf{r}%
_{2}\right)  +C~,\;C=\frac{e^{2}}{\kappa\left\vert \mathbf{r}_{1}%
-\mathbf{r}_{2}\right\vert }~,\nonumber\\
&  h^{0}\left(  \mathbf{r}_{i}\right)  =\frac{1}{2m}\left[  \mathbf{p}%
_{i}-\frac{e}{c}\mathbf{A}\left(  \mathbf{r}_{i}\right)  \right]
^{2}+V\left(  \mathbf{r}_{i}\right)  ~,\;i=1,2~,\nonumber\\
&  \mathbf{A}\left(  \mathbf{r}_{i}\right)  =\frac{B_{i}}{2}\left(
-y_{i},x_{i},0\right)  ~,\;V\left(  x,y\right)  =\frac{m\omega_{0}^{2}}%
{2}\left[  \frac{1}{4a^{2}}\left(  x^{2}-a^{2}\right)  ^{2}+y^{2}\right]  ~.
\label{a1}%
\end{align}
In this expression $c$ is the speed of the light, $\kappa$ is the dielectric
constant of the medium, $m$ is the effective mass of the electron (e.g.,
$m=0,067m_{e}$ in GaAs), $\mathbf{r}_{i}$ and $\mathbf{p}_{i}$ are the
position and the momentum of the $i$-th electron, $B_{i}$ is the magnetic
field at dot $i$ and the harmonic potential well $V\left(  \mathbf{r}\right)
$ of frequency $\omega_{0}$ is motivated by experimental results
\cite{TarAuHHK96}. The matrix elements needed to calculate $J$ can be obtained
by adding and subtracting in (\ref{a1}) the harmonic potentials $V_{\pm}$
centered at $x_{i}=\left(  -1\right)  ^{i}a$ for the $i$-th electron,%
\begin{align}
&  H_{orb}=h_{-}^{0}\left(  \mathbf{r}_{1}\right)  +h_{+}^{0}\left(
\mathbf{r}_{2}\right)  +W+C~,\nonumber\\
&  h_{\pm}^{0}\left(  \mathbf{r}_{i}\right)  =\frac{1}{2m}\left[
\mathbf{p}_{i}-\frac{e}{c}\mathbf{A}\left(  \mathbf{r}_{i}\right)  \right]
^{2}+V_{\pm}\left(  \mathbf{r}_{i}\right)  ~,\nonumber\\
&  V_{\pm}\left(  \mathbf{r}_{i}\right)  =\frac{m\omega_{0}^{2}}{2}\left[
\left(  x_{i}\pm a\right)  ^{2}+y_{i}^{2}\right]  ~,~W=\sum_{i=1}^{2}V\left(
\mathbf{r}_{i}\right)  -\left[  V_{-}\left(  \mathbf{r}_{1}\right)
+V_{+}\left(  \mathbf{r}_{2}\right)  \right]  ~, \label{a2}%
\end{align}
and using the ground states functions $\varphi_{\pm}$ centered at
$\mathbf{r}=\left(  \pm a,0\right)  $,%
\begin{align}
&  \varphi_{\pm}\left(  \mathbf{r}\right)  =\sqrt{\frac{m\omega_{\pm}}%
{\pi\hbar}}\exp\left(  \pm\frac{ima\omega_{L_{\pm}}}{\hbar}y\right)
\exp\left(  -\frac{m\omega_{\pm}\left[  \left(  x\mp a\right)  ^{2}%
+y^{2}\right]  }{2\hbar}\right)  ~,\nonumber\\
&  \omega_{\pm}=\sqrt{\omega_{0}^{2}+\omega_{L_{\pm}}^{2}}~,\ \omega
_{L_{-\left(  +\right)  }}=eB_{1\left(  2\right)  }/2mc~, \label{a3}%
\end{align}
which are eigenfunctions of $h_{\pm}^{0}$ with energy $\omega_{\pm}$,%
\[
h_{-\left(  +\right)  }^{0}\left(  \mathbf{r}_{1\left(  2\right)  }\right)
\varphi_{-\left(  +\right)  }\left(  \mathbf{r}_{1\left(  2\right)  }\right)
=\omega_{-\left(  +\right)  }\varphi_{-\left(  +\right)  }\left(
\mathbf{r}_{1\left(  2\right)  }\right)  ~.
\]
With this, the expression for the interaction function becomes%
\begin{align}
J  &  =\frac{2S^{2}}{\left(  1-S^{4}\right)  }\left[  L-\frac{\hbar\omega_{0}%
}{4}\frac{\left(  b_{+}^{2}-b_{-}^{2}\right)  \left(  b_{-}-b_{+}\right)
}{b_{+}b_{-}}\right]  ~,\nonumber\\
L  &  =\left(  \left\langle 12\right\vert C+W\left\vert 12\right\rangle
-\frac{\operatorname{Re}\left\langle 12\right\vert C+W\left\vert
21\right\rangle }{S^{2}}\right)  ~,\ b_{\pm}=\omega_{\pm}/\omega_{0}~,
\label{a4}%
\end{align}
using the expression (\ref{a3}) for the functions $\varphi_{\pm}\left(
\mathbf{r}_{i}\right)  =\left\langle \mathbf{r}_{i}\right.  \left\vert
\pm\right\rangle $ we get%
\begin{align*}
&  \frac{S^{2}}{\left(  1-S^{4}\right)  }=\frac{1-\Delta^{2}}{\left(
2\sinh\left(  2M\right)  +\Delta\exp\left(  -2M\right)  \left(  2-\Delta
^{3}\right)  \right)  }~,\\
&  M=\frac{2d^{2}}{b_{+}+b_{-}}\left[  b_{-}b_{+}+\frac{\left(  \omega_{L_{+}%
}+\omega_{L_{-}}\right)  ^{2}}{4\omega_{0}^{2}}\right]  ~,\ \Delta=\frac
{b_{-}-b_{+}}{b_{-}+b_{+}}~,\ d=\frac{a}{a_{0}}~,
\end{align*}
where $a_{0}=\sqrt{\hbar/m\omega_{0}}$ is the effective Bohr radius of the
dot. For the matrix elements of $W$ in (\ref{a4}) we have%
\begin{align*}
&  \left\langle 12\right\vert W\left\vert 12\right\rangle -\frac{\left\langle
12\right\vert W\left\vert 21\right\rangle }{S^{2}}=\\
&  \frac{\hbar\omega_{0}}{2}\left\{  \frac{3}{2d^{2}\left(  b_{-}%
+b_{+}\right)  ^{2}}\left[  \frac{1+\Delta^{2}}{\left(  1-\Delta^{2}\right)
^{2}}-1\right]  -3\left(  \frac{\Delta^{2}-1}{b_{-}+b_{+}}\right)
-\frac{d^{2}}{2}\left(  \Delta^{4}-6\Delta^{2}-3\right)  \right\}
\end{align*}
and for the matrix elements of the electric interaction $C$ between the
electrons%
\begin{align*}
&  \left\langle 12\right\vert C\left\vert 12\right\rangle -\frac
{\operatorname{Re}\left\langle 12\right\vert C\left\vert 21\right\rangle
}{S^{2}}=\\
&  \frac{e^{2}}{a_{0}\kappa}\sqrt{\frac{\pi}{2}\bar{b}}\left\{  \sqrt{\left(
1-\Delta^{2}\right)  }\exp\left[  -d^{2}\left(  1-\Delta^{2}\right)  \bar
{b}\right]  \mathrm{I}_{0}\left[  d^{2}\left(  1-\Delta^{2}\right)  \bar
{b}\right]  -\exp\left[  \frac{d^{2}}{2}K\right]  \mathrm{I}_{0}\left[
\frac{d^{2}}{2}K\right]  \right\}  ~,\\
&  K=\bar{b}\left(  1+\Delta^{2}\right)  -\frac{1}{\bar{b}}+\sqrt{\left[
\left(  1-\Delta^{2}\right)  \bar{b}\right]  ^{2}-2\left(  1+\Delta
^{2}\right)  +\frac{1}{\bar{b}^{2}}}~,\ \bar{b}=\frac{b_{-}+b_{+}}{2}~,
\end{align*}
where $\mathrm{I}_{0}$ is the zeroth-order Bessel function. The quantity
$\Delta$ is related to the difference and to the sum of the fields ($B_{\pm
}^{\prime}=B_{1}\pm B_{2}$, without the $g$-factor)\ at the dots by%
\[
\Delta=\frac{B_{+}^{\prime}B_{-}^{\prime}}{2\left[  \left(  2\hbar\frac
{\omega_{0}}{\mu_{B}}\right)  ^{2}+B_{+}^{\prime~2}+B_{-}^{\prime~2}%
+\sqrt{\left[  \left(  2\hbar\frac{\omega_{0}}{\mu_{B}}\right)  ^{2}%
+B_{+}^{\prime~2}+B_{-}^{\prime~2}\right]  ^{2}-\left(  2B_{+}^{\prime}%
B_{-}^{\prime}\right)  ^{2}}\right]  }~.
\]
This quantity can be used to evaluate the error in considering $J$ independent
of $B_{-}^{\prime}$. In the denominator of the above expression, usually only
the first term is relevant, e.g., for a typical GaAs dot we have $\hbar
\omega_{0}/\mu_{B}=50~\mathrm{T}$. For the same field at both dots we obtain
the result given in \cite{BurLoD99}.

\section{The exclusive OR gate\label{XOR}}

Here we establish the parameters of two kinds of parallel external fields that
can act as a XOR gate, that is, fields in the form (\ref{5}) whose evolution
operator (\ref{4.1}), at a given instant of time, has the form (\ref{4a}). The
same analysis presented here can also be carried out for the 26 different
families of external fields presented in \cite{BagGiBL05}, which gives a wide
range of fields to choose from that are more appropriate to adjust to
experimental setups. All the development depends on the ability to vary
independently the difference $B_{-}$ and the interaction $J$, but, as we saw
in the previous sections, there are various circumstances when this
requirement can be satisfied.

\subsection{$B_{-}$ proportional to the interaction}

Suppose that we obtain, theoretically or experimentally, an expression for the
interaction function $J$,\ and we choose $B_{-}$ such that it is proportional
to $J$. As a consequence we can write%
\[
J\left(  t\right)  =q\left(  t\right)  \sin\lambda\,,\;B_{-}\left(  t\right)
=q\left(  t\right)  \cos\lambda\,,
\]
where $q\left(  t\right)  $ is an arbitrary function of time, and $\lambda$ is
a real constant. Then, the general solution of equation (\ref{3e}) can be
written as $\psi\left(  t\right)  =\hat{u}_{t}\psi\left(  0\right)  $ with the
evolution operator $\hat{u}_{t}$ given by%
\begin{equation}
\hat{u}_{t}=\cos\omega-i\left(  \sigma_{1}\sin\lambda+\sigma_{3}\cos
\lambda\right)  \sin\omega\,,\;\omega\left(  t\right)  =\int_{0}^{t}q\left(
\tau\right)  ~d\tau~. \label{p2}%
\end{equation}
Selecting the instants $T$ when $\cos\omega=1$,
\begin{equation}
\omega\left(  T\right)  =\int_{0}^{T}q\left(  \tau\right)  ~d\tau
=2n\pi~,\;n\in\mathbb{N}^{\ast}~, \label{p3}%
\end{equation}
the evolution operator (\ref{4.1}) of this problem assumes the form%
\[
R_{T}=\exp\left(  i3n\pi\right)  \exp\left[  -\frac{i}{2}\left(  \Gamma\left[
\Sigma_{3}+\rho_{3}\right]  +2n\pi\Sigma_{3}\rho_{3}\sin\lambda\right)
\right]  ~.
\]
So to obtain the XOR gate (\ref{4a}), up to a phase, at every instant $T$
(\ref{p3}), we have to set%
\[
\Gamma\left(  T\right)  =2n\pi\sin\lambda=\frac{\pi}{2}\operatorname{mod}%
\left(  2\pi\right)  ~,
\]
or, in a more explicit form,%
\[
\sin\lambda=\frac{4m+1}{4n}~,\ m\in\mathbb{N}~,\ m<n~.
\]

When the interaction $J$ and the difference $B_{-}$ are constants, so $q$ is a
constant, we obtain%
\begin{equation}
\Gamma\left(  T\right)  =JT\operatorname{mod}\left(  2\pi\right)
\,,\;T=\frac{2n\pi}{\sqrt{J^{2}+B_{-}^{2}}}~. \label{const}%
\end{equation}
For the special case of a constant sum of fields $B_{+}$, the above expression
can be compared with the XOR gate obtained in \cite{BurLoDS99}. To have a idea
about the physical values involved in the expression (\ref{const}), for a
typical GaAs QD \cite{BurLoD99}, for an interaction about $50~\mathrm{\mu eV}$
and a $B_{-}$ of $10~\mathrm{mT}$\ we obtain a XOR gate in a time of
$10~\mathrm{ps}$.

\subsection{Adiabatic pulse}

The implementation of the quantum gate is performed by the variation of the
external fields. However, as described in condition (1) of section \ref{4LS},
these variations can not be so fast so as to prevent the excitation of higher
energy-levels. This problem can be avoided by using an adiabatic variation of
the fields, which can be obtained by a time-dependence in the form of
$\mathrm{sech}\left(  \omega t\right)  $ \cite{BurLoDS99}, with $\omega
<<\Delta E/\hbar$ (see condition (1) in section \ref{4LS}). In this section we
analyze the case when the field's difference is constant and the interaction
varies adiabatically. The case when both the interaction $J$\ and the
difference $B_{-}$\ vary adiabatically is a special case of the preceding
section. The case when the interaction function is maintained constant and the
quantity $B_{-}$ vary adiabatically can be study by knowing that
\cite{BagGiBL05}: if $\psi$ is a solution of the equation (\ref{3e}) with the
external field $\left(  K_{1},0,K_{3}\right)  $, then $\tilde{\psi}=\left(
2\right)  ^{-1/2}\left(  \sigma_{1}+\sigma_{3}\right)  \psi\ $is a solution of
(\ref{3e}) with the external field $\left(  K_{3},0,K_{1}\right)  $. In
addition, for $B_{-}$ not proportional to $J$ a lot of solutions of this
$\mathrm{sech}$\ form can be found in \cite{BagGiBL05}.

Suppose that, by vary $B_{+}$ or by controlling the electric potential between
the dots, we obtain a variation in the form%
\begin{equation}
J\left(  t\right)  =a/\cosh\omega t\,,\;B_{-}=c~, \label{cosh}%
\end{equation}
where $a$, $c$ are constants. The evolution operator $\hat{u}_{t}$\ of
equation (\ref{3e}) with the above fields is given by%
\begin{equation}
\hat{u}_{t}\left(  t\right)  =\frac{1}{\left\vert G_{2}^{0}\right\vert
^{2}+\left\vert G_{1}^{0}\right\vert ^{2}}\left(
\begin{array}
[c]{cc}%
G_{1}\left(  z\right)  & -G_{2}^{\ast}\left(  z\right) \\
G_{2}\left(  z\right)  & G_{1}^{\ast}\left(  z\right)
\end{array}
\right)  \left(
\begin{array}
[c]{cc}%
G_{1}^{0~\ast} & G_{2}^{0~\ast}\\
-G_{2}^{0} & G_{1}^{0}%
\end{array}
\right)  ~, \label{tlsad}%
\end{equation}
where the $\ast$ indicates complex conjugation and%
\begin{align}
&  G_{1}\left(  z\right)  =\left(  2c-i\omega\right)  z^{-\nu}\left(
1-z\right)  ^{\nu}F\left(  \lambda,-\lambda;\gamma;z\right)  ~,\nonumber\\
&  G_{2}\left(  z\right)  =2az^{-\nu+1/2}\left(  1-z\right)  ^{\nu
+1/2}F\left(  1+\lambda,1-\lambda,\gamma+1;z\right)  ~,\nonumber\\
&  z\left(  t\right)  =\frac{1}{2}\left(  1-\tanh\omega t\right)
~,\ G_{i}^{0}=G_{i}\left(  1/2\right)  ~,\nonumber\\
&  \nu=-\frac{ic}{2\omega}~,\ \gamma=1/2-2\nu~,\ \lambda=\frac{\left\vert
a\right\vert }{\omega}~. \label{hyper}%
\end{align}
In these expressions, $F\left(  \alpha,\beta,\gamma,z\right)  $ is the Gauss
hypergeometric function.

If we use the \textrm{sech} variation to create a pulse of the interaction
function, this pulse will be turned off when $t>>1/\omega$. In this time limit
we have $\lim_{\omega t\rightarrow\infty}z=e^{-2\omega t}$ and $F\left(
\alpha,\beta,\gamma,0\right)  =1$. If in this limit, we choose%
\begin{equation}
G_{2}^{0}=aF\left(  1+\lambda,1-\lambda,\gamma+1;\frac{1}{2}\right)  =0~,
\label{adcond}%
\end{equation}
the evolution operator (\ref{tlsad}) assumes the form%
\begin{equation}
\hat{u}_{t}\left(  \omega t\rightarrow\infty\right)  =\exp\left(  -i\sigma
_{3}tc\right)  ~. \label{tlslim}%
\end{equation}
So, following the procedure of the preceding sections, we see that the
evolution operator (\ref{4.1}) will behave as a XOR gate (\ref{4a}) in the
instant $T$, when $\Gamma\left(  T\right)  =\pi/2\operatorname{mod}\left(
2\pi\right)  $ and, for $c\neq0$,%
\begin{equation}
a=\frac{\omega\pi\left(  1+4m\right)  }{4\arctan\left[  \exp\left(  \omega
T\right)  \right]  -\pi}~,\ T=\frac{n\pi}{c}~,\ n,m\in\mathbb{N}^{\ast}~,
\label{cond}%
\end{equation}
or, in the limit $\omega T>>1$, $a\simeq\omega\left(  1+4m\right)  $.

The condition (\ref{adcond}) can be obtained in various ways, for example, in
the special case when $c=0$ (and consequently $\lim_{\omega t\rightarrow
\infty}\hat{u}=I$ in (\ref{tlslim}) and we can abandon the condition
$T=n\pi/c$) we can use the relation \cite{Gra71},%
\begin{equation}
F\left(  1+\lambda,1-\lambda;\frac{3}{2};\frac{1}{2}\right)  =\frac{1}%
{\lambda}\sin\left(  \frac{\lambda\pi}{2}\right)  ~. \label{9.137-30}%
\end{equation}
So, we obtain the desired condition whenever $\left\vert a\right\vert
=2m\omega,~m\in\mathbb{N}^{\ast}$. But, once (\ref{cond}) gives $a$ as an odd
multiple of $\omega$, it seems that the gate can not be obtained with equal
fields at both dots.

\section{Some final remarks}

We have demonstrated that the general solution of the Schrödinger equation for
the system of two interacting spins placed in parallel magnetic fields can be
reduced to the general solution of the corresponding equation for only one
spin in an effective external field. This allows us to use the known exact
solutions of the latter problem to obtain exact solutions for the system of
two spins. In turn, this was used by us to describe a construction of a
universal quantum gate. We have found restrictions on external magnetic fields
(in terms of the effective field for the one-spin system) such that the four
level system can act as a XOR gate. The explicit form of the evolution
operator and the known dependence of the spin interaction with the magnetic
field show that a small difference in the fields applied at each dot (or a
small difference in $g$-factors of these dots) can be explored to control the
quantum gate without changing the spin interaction.

One of the problems in the practical application of the theoretical results in
constructing universal quantum gates is the knowledge of the spin interaction
as a function of external fields. This function can be obtained, in a way
unrelated to the Heisenberg interaction model, using a theory that takes into
account the physical characteristics of the system, as we do here using the
Heitler-London approximation. If from the expression of $J$, obtained from
this more complex model, follows that $J$ does not depend on $B_{-}$ for
$B_{-}\ll B_{+}$, we are free to choose $B_{-}$ as a function of $J\left(
B_{+}\right)  $. By using this expression of $J\left(  B_{+}\right)  $, and
choosing $B_{-}\left(  t\right)  $ to be proportional to $J\left(
B_{+}\right)  $, we can measure the probability transition between the states
$\left\vert \uparrow\downarrow\right\rangle $ and $\left\vert \downarrow
\uparrow\right\rangle $ (the swap operation). According to our results, this
probability is equal to $\sin^{2}\lambda\sin^{2}\omega$. This result can be
used to test the consistence between the $J\left(  B_{+}\right)  $, obtained
by any model, and the Heisenberg interaction model.

\begin{acknowledgement}
M.C.B. thanks FAPESP; D.M.G. thanks FAPESP and CNPq for permanent support.{}
\end{acknowledgement}

\end{document}